# Superconducting properties of high-entropy-alloy tellurides M-Te (M: Ag, In, Cd, Sn, Sb, Pb, Bi) with a NaCl-type structure

Md. Riad Kasem[1], Kazuhisa Hoshi[1], Rajveer Jha[1], Masayoshi Katsuno[1], Aichi Yamashita[1], Yosuke Goto[1], Tatsuma D. Matsuda[1], Yuji Aoki[1], Yoshikazu Mizuguchi[1]*

1. Department of Physics, Tokyo Metropolitan University, 1-1, Minami-osawa, 192-0397 Hachioji, Japan.

Corresponding author: Yoshikazu Mizuguchi (mizugu@tmu.ac.jp)

**Abstract**

High-entropy-alloy-type tellurides M-Te, which contain five different metals of M = Ag, In, Cd, Sn, Sb, Pb, and Bi, were synthesized using high pressure synthesis. Structural characterization revealed that all the obtained samples have a cubic NaCl-type structure. Six samples, namely $AgCdSnSbPbTe_5$, $AgInSnSbPbTe_5$, $AgCdInSnSbTe_5$, $AgCdSnPbBiTe_5$, $AgCdInPbBiTe_5$, and $AgCdInSnBiTe_5$ showed superconductivity. The highest transition temperature ($T_c$) among those samples was 1.4 K for $AgInSnSbPbTe_5$. A sample of $AgCdInSbPbTe_5$ showed a semiconductor-like transport behavior. From the relationship between $T_c$ and lattice constant, it was found that a higher $T_c$ is observed for a telluride with a larger lattice constant.

High entropy alloys (HEAs) are defined as alloys containing five or more elements with concentrations between 5 and 35 at% and have been extensively studied in the field of materials science and engineering [1,2]. Recently, a HEA superconductor $Ta_{34}Nb_{33}Hf_8Zr_{14}Ti_{11}$ with a transition temperatures ($T_c$) of 7.3 K was discovered [3]. Although the mechanisms of observed superconductivity in most HEA superconductors are believed to be conventional phonon-mediated pairing, interests in HEA superconductors have been growing due to their notable features and possibilities for material development [4-9]. For example, high-pressure measurements revealed that the superconductivity states in a HEA superconductor, $(TaNb)_{0.67}(HfZrTi)_{0.33}$ are robust at pressures under 190 GPa [10]. This suggests that HEA superconductors could be useful under extreme conditions like high pressure.

In our recent study, we synthesized a HEA-type telluride $AgInSnPbBiTe_5$ where five metals except for Te occupy a single site. $AgInSnPbBiTe_5$ is a superconductor with a $T_c$ of 2.6 K [11]. Since the structure of this phase is cubic NaCl-type (#225, *Fm*-3*m*, $O_h^5$), the structure can be regarded pseudo-binary alloy type with a cationic metal site and anionic tellurium site. Studying HEA-type superconductors is important to develop material design strategies for superconductors with high performance. In fact, we have observed that a layered superconductor $REO_{0.5}F_{0.5}BiS_2$ (RE: rare earth) shows an improvement of superconducting properties by an increase in mixing entropy at the RE site [12,13]. The improvement has been explained by the modification of local crystal structure of the superconducting layer.

Based on those facts, further investigation on the relationship between crystal structure, constituent elements, configurational entropy, effect of disorder, and physical properties in HEA-type compounds is important. Since HEA-type telluride is a simple example of HEA-type compound and shows superconductivity ($AgInSnPbBiTe_5$ [11]), it is a good system to address those issues if various HEA-type tellurides could be obtained. In this study, we decided to investigate similar HEA tellurides $AgCdSnSbPbTe_5$ (#1), $AgCdInSbPbTe_5$ (#2), $AgInSnSbPbTe_5$ (#3), $AgCdInSnSbTe_5$ (#4), $AgCdSnPbBiTe_5$ (#5), $AgCdInPbBiTe_5$ (#6), and $AgCdInSnBiTe_5$ (#7). Those samples were labeled #1–#7 for clarity. According to the charge neutrality in a NaCl-type telluride, the valence state of the metal site should be +2. Basically, the valence state of Sb and Bi tend to be +3 in a NaCl-type structure [14,15]; hence, $Sb^{3+}$ and $Bi^{3+}$ should be incorporated in the tellurides with $Ag^+$ to make the average valence +2. Therefore, we determined the composition to be studied with a restriction that five metals are included, and all the phases contain Ag-Bi or Ag-Sb to satisfy the HEA condition at the metal site.

Precursor powders of HEA-type tellurides were prepared by melting Ag (99.9%), Cd powders (99.9%), In (99.99%), Sn (99.999%), Pb (99.9%), Bi (99.999%), and Te (99.999%) grains with a nominal composition at 800 °C in an evacuated quartz tube. The obtained sample was ground and mixed well by a mortar and pestle. The precursor powders were pelletized into a diameter size of 5 mm and placed in a high-pressure synthesis cell, which was composed of a BN capsule, a carbon capsule (heater), metal electrodes, and a pyrophyllite block. High-pressure annealing was performed at 500°C for 30 min. using a cubic-anvil-type 180-ton press (CT factory) under 3 GPa. The chemical composition of the samples was examined by energy-dispersive X-ray spectroscopy (EDX) on a scanning electron microscope TM-3030 (Hitachi Hightech) equipped with a SwiftED analyzer (Oxford). Five points were chosen on a samples' surface to perform the EDX analysis with a typical analysis area of 30μm × 30 μm. The standard deviation represents the analysis error of the chemical composition. Powder X-ray diffraction (XRD) was performed on a MiniFlex-600 diffractometer (RIGAKU) with a Cu-Kα radiation and a D/teX-Ultra detector by a conventional $\theta$–$2\theta$ method. The resulting XRD patterns were refined by the Rietveld analysis using a RIETAN-FP software [16]. The Rietveld refinements were performed with fixed occupation parameters for both cation and anion sites according to the chemical composition of the samples. Isotropic displacement parameters were also fixed. For the investigation of superconducting properties, temperature dependencies of electrical resistivity were measured on PPMS (Physical Property Measurement System, Quantum Design) using a $^3$He probe by a conventional four-probe method. Au wires (diameter in 25 μm) were attached on the sample surface using Ag paste.

The compositions of the prepared samples analyzed using EDX were summarized in Table I. The obtained compositions are close to the starting nominal compositions. The population range of the elements at the M site is 14–25% among all the samples, which satisfies the HEA criterion of 5-35 at% for the M site. The composition of the Te site is close to 1, which is consistent with the assumption that the examined materials have a NaCl-type structure.

Figure 1(a) shows the XRD patterns for all the obtained samples. The XRD peaks of the major phase can be indexed by a cubic NaCl-type structural model. The lattice constant was estimated using Rietveld refinement and summarized in Table I. According to the elemental configuration, the lattice constant differs, and the shift in XRD peak is observed as shown in Fig. 1(b). Figure 1(c) shows a typical Rietveld refinement result on the XRD profile for sample #3.

Figure 2(a) shows the temperature dependences of the electrical resistivity ($\rho$) for #1, #3, #4, #5, #6, and #7. The temperature dependence of resistivity for #1 and #7 is metallic, but the residual resistivity ratio is very small, which is a behavior similar to conventional HEA superconductors [3-11]. In addition, some samples show almost temperature-independent or weakly-localized behavior in the temperature dependence of resistivity; a similar behavior was observed In-doped PbTe [17]. Apart from the different transport trends, all those samples show a superconducting transition. Figure 2(b) shows the temperature dependences of resistivity at low temperatures below 2 K. Samples #1, #3, #4, #6, and #7 show zero-resistivity states. Sample #5 shows a broad transition, and only onset of superconductivity is observed within the temprature range in this study. The highest Tc was observed for #3: the onset temperature ($T_c^{onset}$) is 1.4 K, and the zero-resistivity temperature ($T_c^{zero}$) is 1.3 K. All the estimated $T_c$ is summarized in Table. I.

Figure 3(a) shows the temperature dependence of resistivity for sample #2, which shows a semiconducting behavior. The Arrhenius plot (log$\rho$-$T^{-1}$ plot) is shown in Fig. 3(b). For all the temperature range, the resistivity data could not be fit with a conventional semiconducting character. Also, we have tested to fit the resistivity data with variable hopping models, but all the tested models could not be suitable for the obtained data.

We discuss the relationship between the lattice constant and the physical properties. First, $T_c^{zero}$ is plotted as a function of lattice constant in Fig. 4. $T_c$ tends to increase with increasing lattice constant. To compare with other metal tellurides, $T_c$-lattice constant data for typical telluride superconductors with a NaCl-type structure [17–23] was plotted in Fig. 4. Here, we choose doped post-transition-metal tellurides; those with small carrier concentration were excluded from this plot. From all the data, we notice a general trend that a $T_c$ is higher for a sample with a larger lattice constant. Although the band structure and superconductivity characteristics for those tellurides are different, such a general trend may be useful to understand the characteristics of highly disordered HRA-type tellurides. As compared to those with one or two metals in the M site, the HEA tellurides have a relatively lower $T_c$. This trend may indicate that high mixing entropy is not favorable for superconductivity in a metal telluride system with a NaCl-type structure. However, although $T_c$ of HEA tellurides is not high, successful synthesis of HEA telluride superconductors and the relationship between lattice constant and $T_c$ revealed here will be useful for further development of various kinds of HEA-type functional materials. We briefly discuss about the observed semiconducting behavior in sample #2. We synthesized two samples for the composition of #2 and observed similar semiconducting behavior for both samples. The lattice constant for sample #2 is 6.150 Å, which is almost middle value among all the samples. Therefore, from structural point of view, we cannot explain the observed semiconducting behavior in sample #2. One possibility is the presence of large disorder in the compound. From the XRD peak structure, we notice that the 200 peak of sample #2 in Fig. 1(b) has a shoulder structure in low angles, which indicates the presence of a phase with a larger lattice constant. Although EDX analyses did not observe clear phase separation, there may be local phase separation (or compositional fluctuations). Those local phase separation may have caused the semiconducting behavior in the sample. Similar XRD result was observed for sample #4 with a low $T_c^{onset}$ of 0.7 K. The presence of local phase separation may affect superconducting properties of HEA-type tellurides as well. However, to understand the effects of the presence of local phase separation or compositional fluctuations to physical properties of HEA tellurides, we need to investigate electronic states of the HEA tellurides, which is challenging issue due to the random occupation of the elements.

In conclusion, we have synthesized HEA tellurides M-Te, which contains five different metals of M = Ag, In, Cd, Sn, Sb, Pb, and Bi using high pressure synthesis. Structural characterization revealed that all the obtained samples have a cubic NaCl-type structure. Six samples, namely $AgCdSnSbPbTe_5$ (#1), $AgInSnSbPbTe_5$ (#3), $AgCdInSnSbTe_5$ (#4), $AgCdSnPbBiTe_5$ (#5), $AgCdInPbBiTe_5$ (#6), and $AgCdInSnBiTe_5$ (#7) showed superconductivity. The highest $T_c$ among those samples was 1.4 K for #3. $AgCdInSbPbTe_5$ (#2) showed a semiconductor-like transport behavior. From the relationship between $T_c$ and lattice constant, it was found that a higher $T_c$ is observed for a telluride with a larger lattice constant. $T_c$ of HEA tellurides is relatively lower than those of conventional metal tellurides with one or two metal elements in the M site.

**Acknowledgements**

This work was partly supported by grants in Aid for Scientific Research (KAKENHI) (Grant Nos. 15H05886, 15H05884, 16H04493, 18KK0076, and 15H03693) and the Advanced Research Program under the Human Resources Funds of Tokyo (Grant Number: H31-1).

Table:1: Composition analyzed by EDX, lattice constant $a$, reliability factor $R_{wp}$ of Rietveld refinement, and transition tempratures ($T_c^{onset}$ and $T_c^{zero}$) for all the synthesized samples.

| Label | Analyzed composition | $a$ (Å) | $R_{wp}$ (%) | $T_c^{onset}$ (K) | $T_c^{zero}$ (K) |
|---|---|---|---|---|---|
| #1 | $Ag_{0.20(1)}Cd_{0.20(2)}Sn_{0.20(2)}Sb_{0.15(2)}Pb_{0.20(1)}Te_{1.05(2)}$ | 6.186(5) | 10.2 | 1.2 | 0.9 |
| #2 | $Ag_{0.24(1)}Cd_{0.17(1)}In_{0.23(2)}Sb_{0.16(2)}Pb_{0.18(1)}Te_{1.02(2)}$ | 6.150(10) | 15.2 | - | - |
| #3 | $Ag_{0.24(1)}In_{0.22(1)}Sn_{0.18(1)}Sb_{0.14(1)}Pb_{0.19(1)}Te_{1.03(1)}$ | 6.202(8) | 9.9 | 1.4 | 1.3 |
| #4 | $Ag_{0.22(2)}Cd_{0.22(2)}In_{0.23(2)}Sn_{0.17(3)}Sb_{0.14(2)}Te_{1.02(1)}$ | 6.098(9) | 12.1 | 0.7 | 0.6 |
| #5 | $Ag_{0.19(1)}Cd_{0.19(1)}Sn_{0.20(2)}Pb_{0.18(1)}Bi_{0.21(1)}Te_{1.03(2)}$ | 6.244(5) | 10.7 | 1.0 | - |
| #6 | $Ag_{0.21(1)}Cd_{0.19(2)}In_{0.25(3)}Pb_{0.16(1)}Bi_{0.18(1)}Te_{1.00(5)}$ | 6.189(13) | 14.5 | 1.0 | 0.9 |
| #7 | $Ag_{0.21(1)}Cd_{0.21(1)}In_{0.24(1)}Sn_{0.19(1)}Bi_{0.19(1)}Te_{0.97(1)}$ | 6.136(10) | 10.0 | 1.0 | 0.9 |

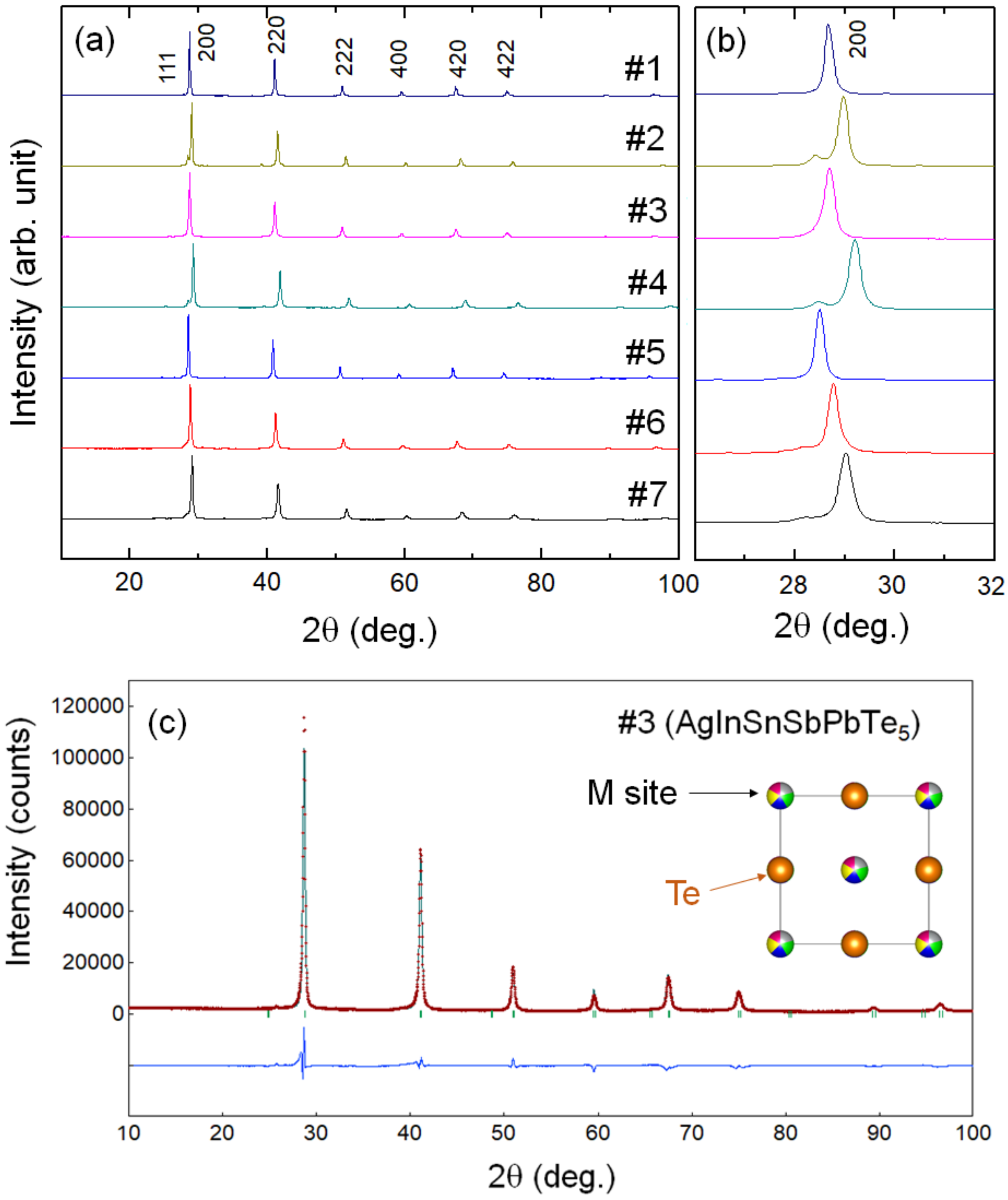

Figure 1. (a) X-ray diffraction (XRD) patterns for all the obtained samples (#1–#7). (b) Zoomed XRD profiles near the 200 peak for all the samples. (c) Typical result of Rietveld refinement (sample #3). Inset shows a schematic image of NaCl-type structure of M-Te.

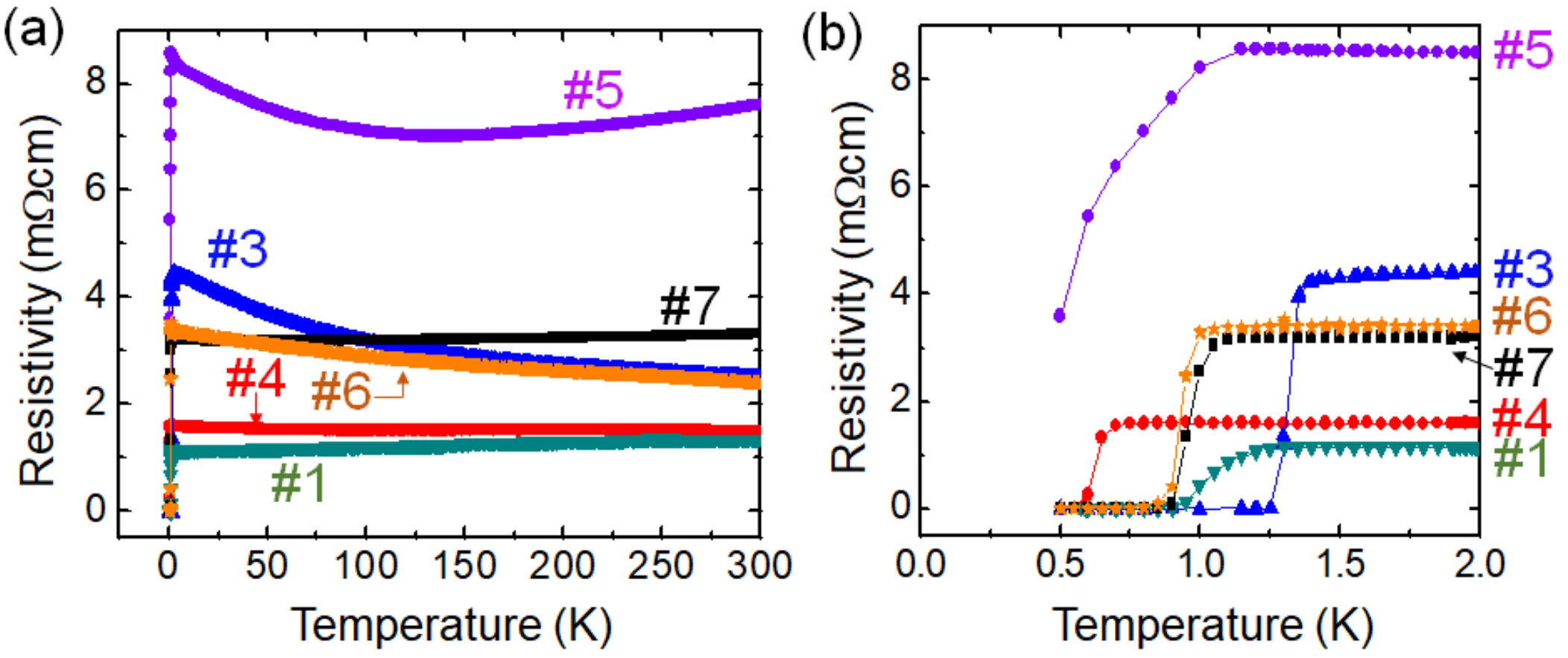

Figure 2. (a) Temperature dependences of electrical resistivity for #1, #3, #4, #5, #6, and #7. (b) Zoomed plots of the temperature dependence of resistivity for #1, #3, #4, #5, #6, and #7 at low temperatures.

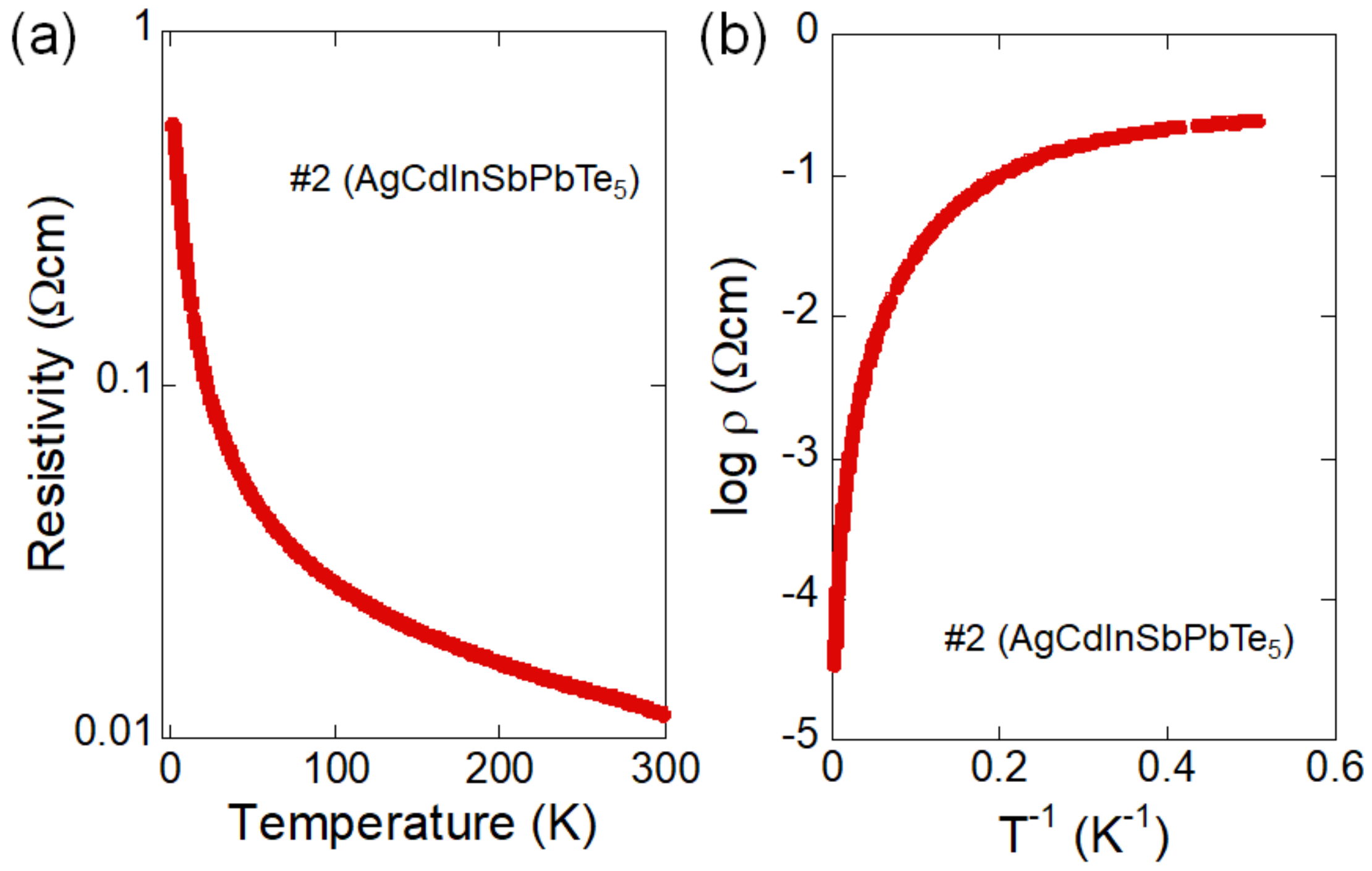

Fig. 3: (a) Temperature dependence of electrical resistivity for sample #2 ($AgCdInSbPbTe_5$). (b) log$\rho$-$T^{-1}$ plot for sample #2.

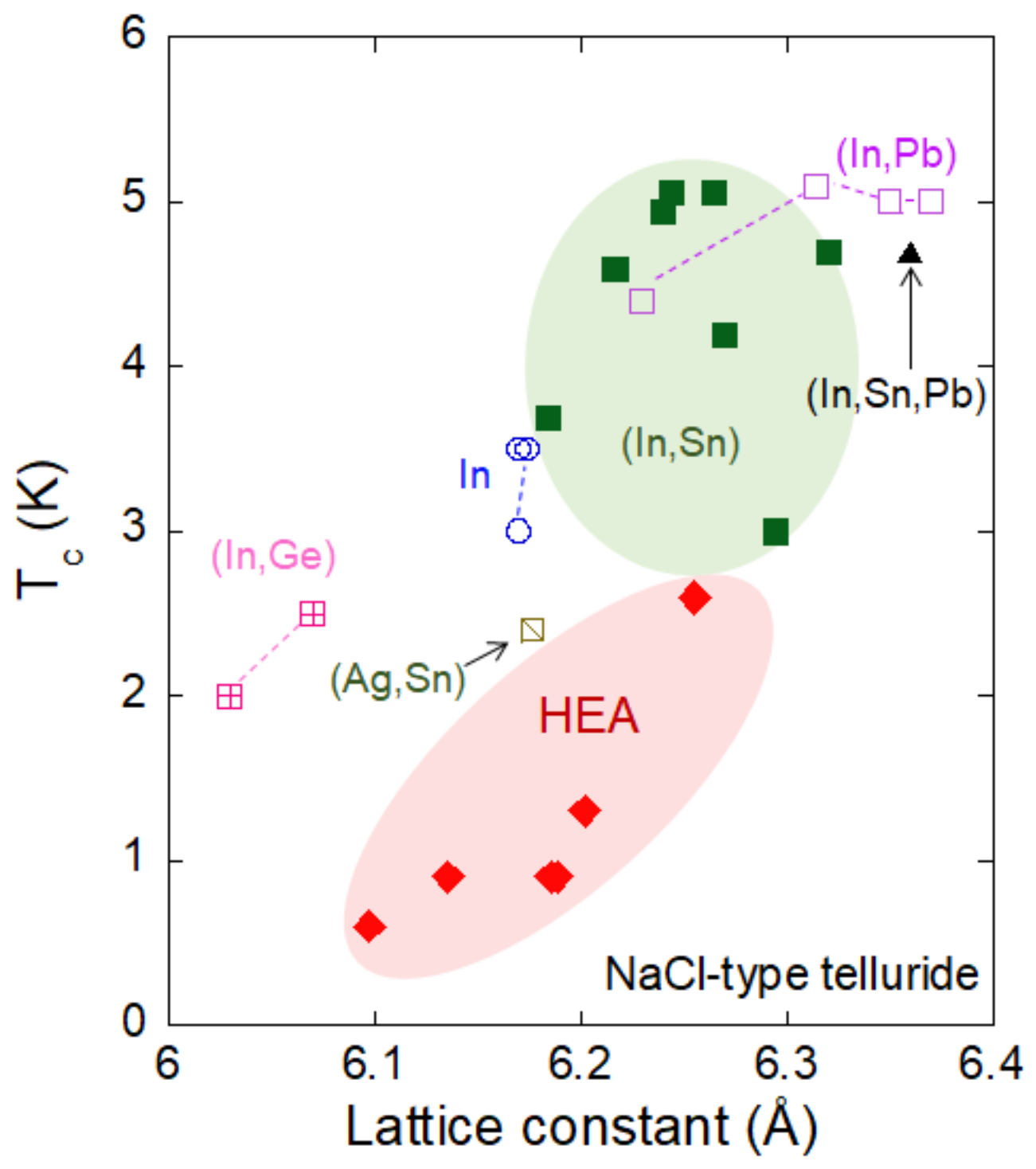

Figure 4. Superconducting transition temperatures ($T_c$) of various metal telluride plotted as a function of lattice constant. The data were taken from Refs. 17–23.